\documentclass[aps,amsmath,amssymb,reprint]{revtex4-1}

\usepackage{graphicx}
\usepackage{dcolumn}
\usepackage{bm}

\usepackage[utf8]{inputenc}

\usepackage[colorlinks,linkcolor=blue,anchorcolor=blue,citecolor=blue]{hyperref}

\begin{document}

\title{\bf New Taub-NUT Black Holes with Massive Spin-2 Hair}%
\author{Yu-Qi Chen}%
\email{yuqi\_chen@tju.edu.cn}
\author{Hai-Shan Liu}
\email{hsliu.zju@gmail.com}

\affiliation{\it Center for Joint Quantum Studies and Department of Physics,\\
School of Science, Tianjin University, Tianjin 300350, China}

\begin{abstract}
We consider Einstein gravity extended with quadratic curvature invariants, where the well-known Ricci-flat Taub-NUT black hole remains a solution. An analysis of the unstable Lichnerowicz modes in the Taub-NUT background enables us to identify the mass and NUT parameters $(m,n)$ where new Taub-NUT black holes can emerge. We then adopt numerical technique to construct these new Taub-NUT black holes that bifurcate away from the Ricci-flat ones. Unlike the Ricci-flat Taub-NUT, there can exist two new black holes for an appropriate given temperature, making it a total of three if we include the Ricci-flat one.
\end{abstract}

\maketitle

\clearpage

\textit{Introduction.}- In the beginning of last century, Einstein discovered General Relativity (GR), which has two important predictions. One is the black hole and the other is gravitational wave. In this century, people have directly detected gravitational waves \cite{LIGOScientific:2017vwq,LIGOScientific:2017zic} and taken images of black holes\cite{EventHorizonTelescope:2019dse,EventHorizonTelescope:2019uob,EventHorizonTelescope:2019jan,EventHorizonTelescope:2019ths,EventHorizonTelescope:2019pgp,EventHorizonTelescope:2019ggy}. These great successes validate GR as a classical theory. However, GR is not a perfect theory; it is not renormalizable and can have have an infinite possibility of higher order corrections \cite{tHooft:1974toh}. It was pointed out that one can obtain a renormalizable theory by including also the quadratic curvature terms to the Einstein-Hilbert action. The price to pay is that these higher-derivative terms that make the theory renormalizable will inevitably introduce ghost modes \cite{Stelle:1976gc}. Many works have been done on addressing the ghost phenomena of higher derivative gravity \cite{Horava:2009uw,Lu:2009em,Li:2008dq, Lu:2010ct,Lu:2010cg,Lu:2011zk} and it was pointed out that the ghost problem might not be severe to the theory \cite{Smilga:2013vba}.

There are three quadratic curvature terms in general, Ricci scalar square, Ricci tensor square and Riemann tensor square($R^2, R_{\mu\nu}R^{\mu\nu}, R_{\mu\nu\rho\sigma}R^{\mu\nu\rho\sigma}$). We can regroup them as the Gauss-Bonnet combination $R^2 + 4 R_{\mu\nu}R^{\mu\nu} + R_{\mu\nu\rho\sigma} R^{\mu\nu\rho\sigma}$, Weyl-squared term $C^{\mu\nu\rho\sigma} C_{\mu\nu\rho\sigma}$ and $R^2$. In four dimensions, the Gauss-Bonnet term is a total derivative and hence doesn't contribute to the equations of motion. The remaining Weyl-squared term and $R^2$ will introduce massive spin-2 and massive scalar modes respectively. The absence of the massive scalar mode can lead to critical gravity\cite{Lu:2011zk}.  One important property of the quadratic curvature extension in four dimensions is that Ricci-flat metrics continue to be the solutions of the extended theory. This raises a question: does the theory admit new black holes beyond the Schwarzschild black hole?  It turns out that the Ricci scalar must be zero owing to the application of the no scalar theorem in higher-derivative gravity \cite{Lu:2015cqa}. On the other hand, black holes beyond the Schwarzschild black hole can carry massive spin-2 hair and they were constructed in \cite{Lu:2015cqa}.

Taub-NUT spacetime is another one of the simplest Ricci-flat solutions in Einstein gravity which was constructed in 1960s \cite{Taub:1950ez,Newman:1963yy}. Compared to the famous Schwarszchild black hole, Taub-NUT spacetime has an additional integration constant $n$, which is called NUT parameter. Due to this NUT parameter, it has a string-like singularity along the polar axis, which is now known as Misner string singularity\cite{Misner:1963fr}. Surprisingly, the Taub-NUT metric has no curvature singularity though it  possesses an event horizon and two other Killing horizons. Those peculiar properties of Taub-NUT spacetime have been attracting  many attentions for years \cite{Clement:2015cxa,Clement:2015aka,Hennigar:2019ive,Bordo:2019slw,Durka:2019ajz,Awad:2022jgn,Wu:2019pzr,Rodriguez:2021hks,Liu:2022wku,Liu:2023uqf,Jiang:2019yzs,Chen:2023eio,Siahaan:2022jrl,Yang:2023hll}.

It is clear the Ricci-flat Taub-NUT black hole will continue to be the solution in the extended gravity theory. In this paper, we would like to investigate whether new black holes carrying the NUT parameter beyond the Ricci-flat solution can arise. As in the case of \cite{Lu:2015cqa}, the no-hair theory will rule out the possibility of having the hair associated with the massive scalar mode. We focus on the construction involving the massive spin-2 mode. The general solution should contain two parameters $(m,n)$. In the Ricci-flat case, any $(m,n\ne 0)$ can lead to an event horizon, but this is no longer true when higher-derivative terms are involved. The parameters $(m,n)$ has to be fine tuned in order to form an event horizon. This fine-tuning leads to difficulties in constructing the numerical solutions, since we have to start at some fiducial point of the parameter space.

Since the massive spin-2 mode of the quadratic extension is relevant for the new solution, we perform a spin-2 tensor mode analysis upon the Ricci-flat Taub-NUT background, which is called Lichnerowicz mode analysis. We find that there exists a Licnerowicz mode for some specific values of $(m,n)$ with a negative eigenvalue, which implies that the Taub-NUT background of this $(m,n)$ is not stable and a more stable and new black hole solution should emerge at this point of the parameter space. With this guidance we construct new Taub-NUT like black hole solutions carrying spin-2 hair in the theory of Einstein gravity added with quadratic curvature terms.

\textit{Theroy.}- We consider Einstein gravity extended with the most general quadratic terms in four dimensional spacetime
\begin{equation} \label{theory}
    I=\int d^{4}x\sqrt{-g}(\kappa R-\alpha C_{\mu\nu\rho\sigma}C^{\mu\nu\rho\sigma}+\beta R^2) \,,
\end{equation}
where $C_{\mu\nu\rho\sigma}$ is Weyl tensor and $\kappa,~\alpha$ and $\beta$ are constants.  For simplicity we shall set $\kappa = 1$. The equations of motion can be obtained through the variation of the metric and  written in a simple form
\begin{equation} \label{equation}
\begin{split}
        &R_{\mu\nu}-\frac{1}{2}g_{\mu\nu}R-4\alpha B_{\mu\nu}
    +2\beta (R_{\mu\nu}\\&-\frac{1}{2}g_{\mu\nu}R+g_{\mu\nu}\nabla_{\rho}\nabla^{\rho}R-\nabla_{\mu}\nabla_{\nu}R)=0
\end{split}
\end{equation}
where $B_{\mu\nu}=(\nabla^{\rho}\nabla^{\sigma}+\frac{1}{2}R^{\rho\sigma})C_{\mu\nu\rho\sigma}$ is the trace free Bach tensor.

 It is worth pointing out that we don't use the Gauss-Bonnet terms to get rid off the Riemann square but to group the quadratic curvature terms into Weyl square. The advantage is that the Einstein gravity theory with quadratic curvature terms  contains three modes in the linear spectrum, while the three terms in the theory (\ref{theory}) correspond to the three modes respectively,  a usual massless spin-2 graviton, a massive spin-2 mode with mass $m_{2}= 1/\sqrt{2\alpha}$ and a massive scalar mode with mass $m_{0}=1/\sqrt{6\beta}$.  It was proved that the scalar hair is forbidden for static black hole solution\cite{Lu:2015cqa}, since the trace part of the equation of motion leads to the vanishing of the Ricci scalar, $R=0$. However, the massive spin-2 mode can survive, and the theory can admit black hole solutions carrying spin-2 hair . In this paper, we want to explore the situation in Taub-NUT spacetimes.

The Taub-NUT solution was constructed in pure Einstein gravity,

\begin{equation}\label{taub}
    ds^2_4 = - h(dt + 2 n \cos \theta d\phi) ^2 + \frac{dr^2}{f} + (r^2 + n^2) d\Omega^2 \,,
\end{equation}

with
\begin{equation}
    h=f = \frac{r^2 - 2 m r - n^2}{r^2 + n^2}
\end{equation}

The solution is a generalization of the famous Schwarzschild black hole solution. In add to the the mass parameter $m$, the Taub-NUT has one more integration constant $n$ which is usually called NUT parameter. Setting $n=0$, the Taub-NUT solution reduces to Schwarzschild solution.

 As is mensioned, the Taub-NUT solution is naturally a solution of theory (\ref{theory}), since the solution is Ricci flat, $R_{\mu\nu} = 0$. It is easily to verify that the Ricci scalar also vanishes under the Taub-NUT like spacetime (\ref{taub}) with arbitrary  $h(r)$ and  $f(r)$ satisfying condition of $h$ and $f$ vanish on the event horizon. The trace of the equations of motion (\ref{equation}) is $\beta (\Box - m_0^2) R = 0$. Multiplying $R$ and integrating over the horizon to infinity gives
$\int dx^4 \sqrt g R (\Box - m_0^2) R =  \int dx^4 \Big[ \partial_r ( \sqrt g g^{rr} R \partial_r R ) - \sqrt g ( g^{rr} \partial_r R \partial_r R + m_0^2 R^2 ) \Big] = 0$. Since $g^{rr}$ vanishes on the horizon and $\partial_r R$ goes to 0 at spatial infinity, the total derivative term gives no contribution, and the remaining two terms are non positive, thus the Ricci scalar has to be 0. Any Taub-NUT like solution of metric ansatz (\ref{taub}), with undetermined $h$ and $f$,   in the theory (\ref{theory}) will have a vanishing Ricci scalar. This property of vanishing Ricci scalar makes a great simplification to the equations of motion.

\textit{Lichnerowicz Mode.}- Since Taub-NUT black hole is a solution of the theory (\ref{theory}), we want to analyze the linear exited modes under the Taub-NUT background. The Taub-NUT spacetime is locally flat, $R_{\mu\nu} = 0$, making a perturbation of Ricci tensor, we find
\begin{equation}\label{pert}
\begin{split}
       \delta R_{\mu\nu} - \frac{1}{2} g_{\mu\nu}\delta R + (2 \beta - \frac{2}{3} \alpha) (g_{\mu\nu} \Box - \nabla_\mu \nabla_\nu ) \delta R
 \\- 2 \alpha \Box ( \delta R_{\mu\nu} - \frac{1}{2} g_{\mu\nu} \delta R ) -4 \alpha R_{\mu\rho\nu\sigma \delta} \delta R^{\rho \sigma} = 0 .
\end{split}
\end{equation}

The trace of above equation is$(6 \beta  \Box  - 1) \delta R = 0$. With the same analysis in the last section, we can deduce that the $\delta R$ must vanish. It means that any Taub-NUT like black hole which is perturbatively around Taub-NUT spacetime must have a vanishing Ricci scalar at the linearised level. Substituting $\delta R = 0$ back to equation (\ref{pert}) gives
\begin{equation}
    (\Delta_L  + \frac{1}{2 \alpha}) \delta R_{\mu\nu} = 0 ,
\end{equation}
where
\begin{equation}
    \Delta_L \delta R_{\mu\nu} = - \Box \delta R_{\mu\nu} - 2 R_{\mu\nu\rho\sigma} \delta R^{\rho\sigma}
\end{equation}
 is called Lichnerowicz operator. For Ricci flat spacetime($R_{\mu\nu} = 0$) such as Taub-NUT black hole, firstly, we can deduce that $g^{\mu\nu} \delta R_{\mu\nu} =0$ from $\delta R = 0$. Secondly, taking variation of the Bianchi identity $\nabla^\mu R_{\mu\nu} - (1/2) \nabla_\nu R = 0$, we can have $\nabla^\mu \delta R_{\mu\nu} = 0$. Thus, $\delta R_{\mu\nu}$ is a transverse traceless tensor on the Ricci flat background.

This implies that once the Lichnerowicz operator has a transverse traceless eigenfunction $\psi_{\mu\nu}$  with eigenvalue $\lambda = -1/(2\alpha)$, $\Delta_L \psi_{\mu\nu} = \lambda \psi_{\mu\nu}$, there exists a linearized perturbation away from the Taub-NUT solution.

In the large $r$ limit, the Lichnerowicz equations can be solved and two Yukawa like modes, $e^{\pm \sqrt{- \lambda r}}$, emerge, which is under expectation.  In general, the Lichnerowicz equation can not be solved analytically, and to solve the equation numerically, one need to use shooting method to seek proper values of the parameters to elude the diverging mode $e^{+\sqrt{\lambda r}}$, left only the normalisable mode $e^{-\sqrt{\lambda r}}$. Without loss of generality, we set $\alpha = 1/2$, and find that there exists one mode corresponding to each value of $n$, and relation between the starting point $r_0$ of the linear perburbation and $n$ are shown in Fig 1. The technical details can be found in the Appendix. We can see that when $n=0$, $r_0 = 0.876$, which covers the result for Schwarszchild case \cite{Lu:2015cqa}.
\begin{figure}
    \centering
    \includegraphics[scale=0.3]{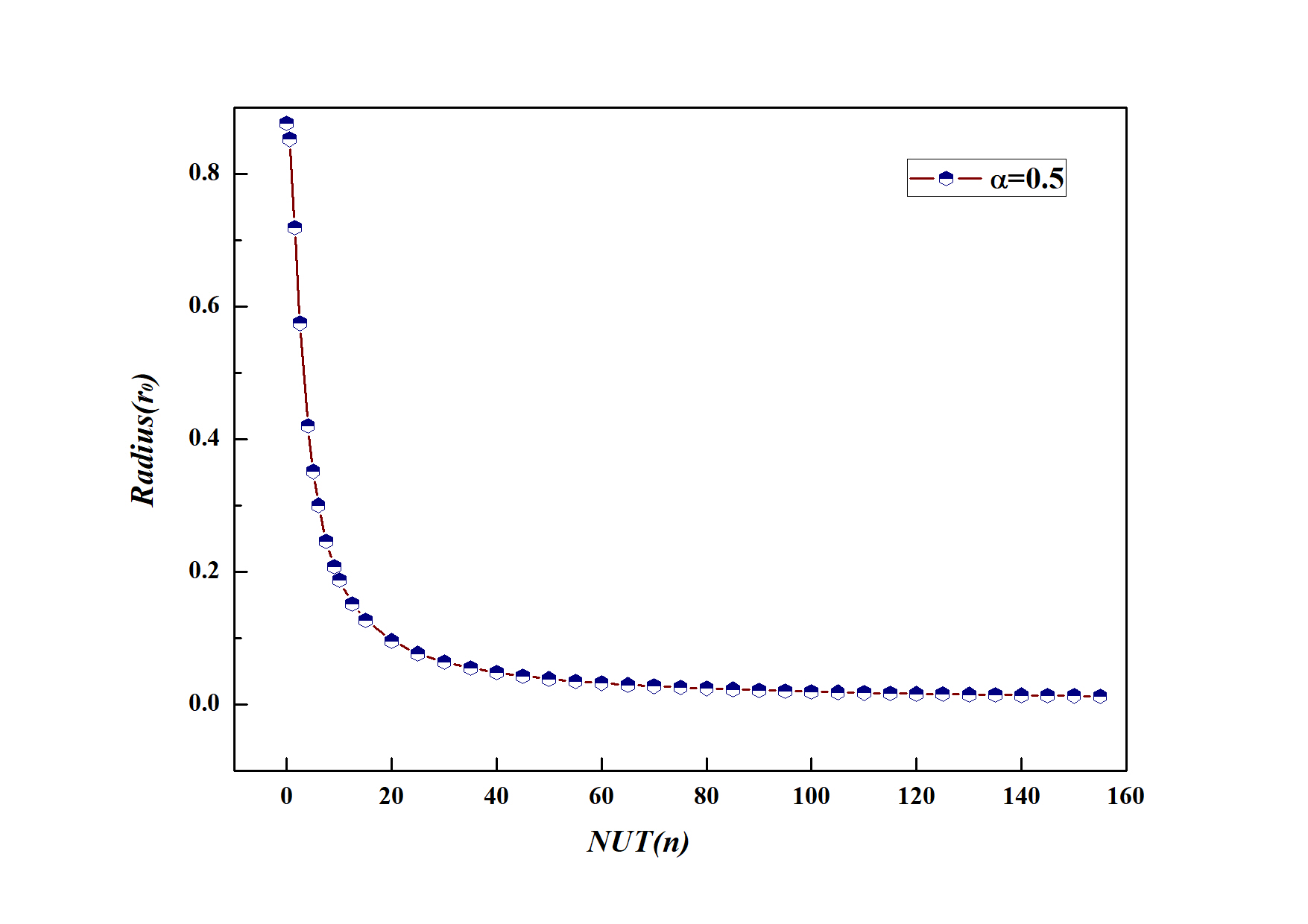}
    \caption{ The bifurcation point ($r_0$,\, n) between linear perturbation and the analytic Taub-NUT solution with the fixed parameter $\alpha=1/2$.}
    \label{fig:my_label}
\end{figure}

\textit{New Taub-NUT Black Hole Solutions.}- The Lichnerowicz mode analysis shows that a spin-2 perturbation exists under the analytic Taub-NUT background, which implies that there must exist black hole solutions carrying spin-2 hair. At this stage, we carry out a numerical search of Taub-NUT like black hole solutions which carry spin-$2$ hair in the theory (\ref{theory}). We start with Taub-NUT like metric ansatz (\ref{taub}) and suppose there exists a black hole with event horizon located at radius $r_0$, where metric functions $h$ and $f$ both vanish. Then, we take a Taylor expansion around the horizon for $h$ and $f$,
\begin{equation}
\begin{split}
        f(r)=f_{1}(r-r_{0})+f_{2}(r-r_{0})^2+f_{3}(r-r_{0})^3+...  \\
    h(r)=h_{1}(r-r_{0})+h_{2}(r-r_{0})^2+h_{3}(r-r_{0})^3+...
\end{split}
\end{equation}

Substituting these back into the equations of motion, the coefficients $h_i$ and $f_i$ for $i>1$ can be solved and expressed in terms of $f_1\,,h_1$ and $r_0$, the first two coefficients are
\begin{equation}
    \begin{split}
        f_{2}=\frac{-3n^2[1+4\alpha f_{1}(f_{1}-h_{1})-f_{1}r_{0}]}{8\alpha f_{1}r_{0}(r_{0}^2+n^{2})}
        \\+\frac{8\alpha f_{1}(1-2f_{1}r_{0})+3r_{0}(f_{1}r_{0}-1)}{8\alpha f_{1}(r_{0}^2+n^{2})}
        \\h_{2}=\frac{h_{1}n^2[1+4\alpha f_{1}(f_{1}-h_{1})-f_{1}r_{0}]}{8\alpha f_{1}r_{0}(r_{0}^2+n^{2})}
        \\+\frac{8\alpha f_{1}(-1+2f_{1}r_{0})+r_{0}(f_{1}r_{0}-1)}{8\alpha f_{1}(r_{0}^2+n^{2})}
    \end{split}
\end{equation}

We use these near-horizon expansion to set initial data and then integrate the solution from horizon to large $r$. Since Taub-NUT solution corresponds to $f_{1}=1/r_{0}=h_{1}$, we set $f_{1}=\frac{1}{r_{0}}+\delta,~h_{1}=\frac{1}{r_{0}}$. And $\delta$ characterizes the deviation of our newly constructed numerical solution from the analytical Taub-NUT solution.
\begin{figure}
    \centering
    \includegraphics[scale=0.14]{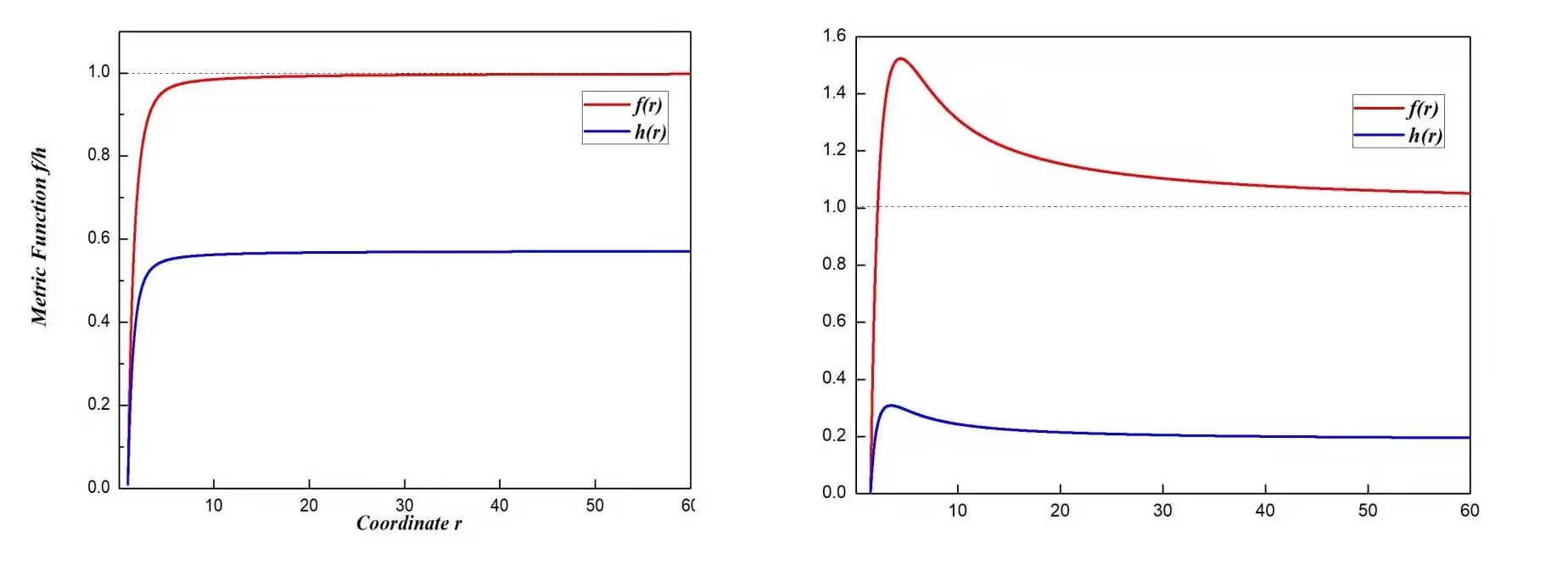}
    \caption{The metric functions $f(r)$ and $h(r)$ for the asymptotically-flat spacetimes with $n=0.5$. The left and right plots show the balk holes with $r_{0}=1$ and $r_{0}=1.5$ respectively.}
    \label{fig:1}
\end{figure}

We set the theory parameter $\alpha = 1/2$ as we did in the last section, and $n=1/2$, then start the investigation of new solution around $r_0 = 0.852$ as suggested by the result of the Lichnerowicz mode analysis. We find a newly Taub-NUT like black holes carrying spin-2 hair under expectation. We show the metric functions $h$ and $f$ in the Fig.2, it is obvious that the two metric functions $h$ and $f$ are different, unlike that of the analytical Taub-NUT solution in which $h$ and $f$ are the same. The profile of metric functions are qualitatively different for different horizon radius. The Fig.2 shows the plot of metric function with $r_0 =1$ and $r_0=1.5$, corresponding to black hole with a positive mass parameter and  black hole with a negative mass parameter, respectively. We read of the mass parameter $m$ from the fall off of metric function $f\sim 1 - 2 m/r+\dots$ in the large $r$. As is for analytic Taub-NUT black hole solution, there is no objection for the mass parameter taking negative value, and the mass of the black hole can have contributions from both mass parameter $m$ and the NUT parameter $n$ \cite{Liu:2022wku,Liu:2023uqf}.  We also show the pattern of the profile of metric function with negative or positive mass parameter for various NUT parameter $n$, whilst keep  $r_0$ fixed in Fig.3.

\begin{figure}
    \centering
    \includegraphics[scale=0.3]{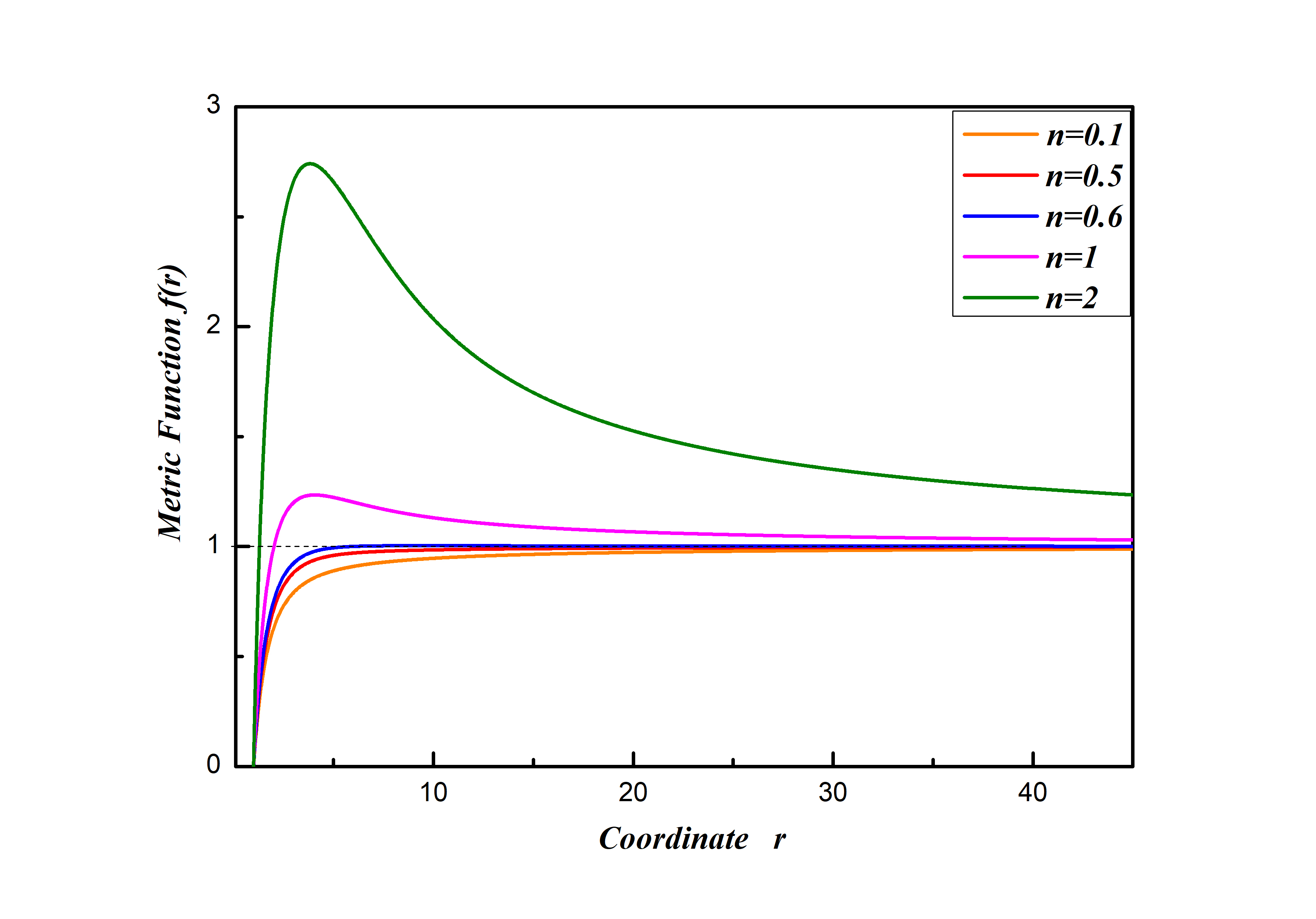}
    \caption{The metric function $f(r)$ with $r_{0}=1$ and variable NUT parameters $n$. The function $f(r)$ with $n=0.1,0.5,0.6,1,2$ are shown as orange, red, blue, pink, and green lines. }
    \label{fig:2}
\end{figure}

In fact, this pattern also emerges in the analytical Taub-NUT solution, where one can easily derive that  the condition for  critical value of mass parameter $m=0$ is $r_0 = n$. Here, for our numerical solution, we can not give an analytical relation of ($r_0\,, n$) for $m = 0$, but we can plot the curve of the relation ($r_0\,, n$) in the $r_0-n$ space, which is displayed in Fig.4. We also plot the cure $r_0 = n$ of analytic Taub-NUT solution as a comparison, the critical curve of the newly found Taub-NUT like solution crosses with the critical curve of analytical Taub-NUT at $ n = 0.83 $. It is worth pointing out that when $n=0$, the critical value of $m=0$ emerges at $r_0= 1.143 $, which is consistent with the result in \cite{Lu:2015cqa}.

\begin{figure}
    \centering
    \includegraphics[scale=0.3]{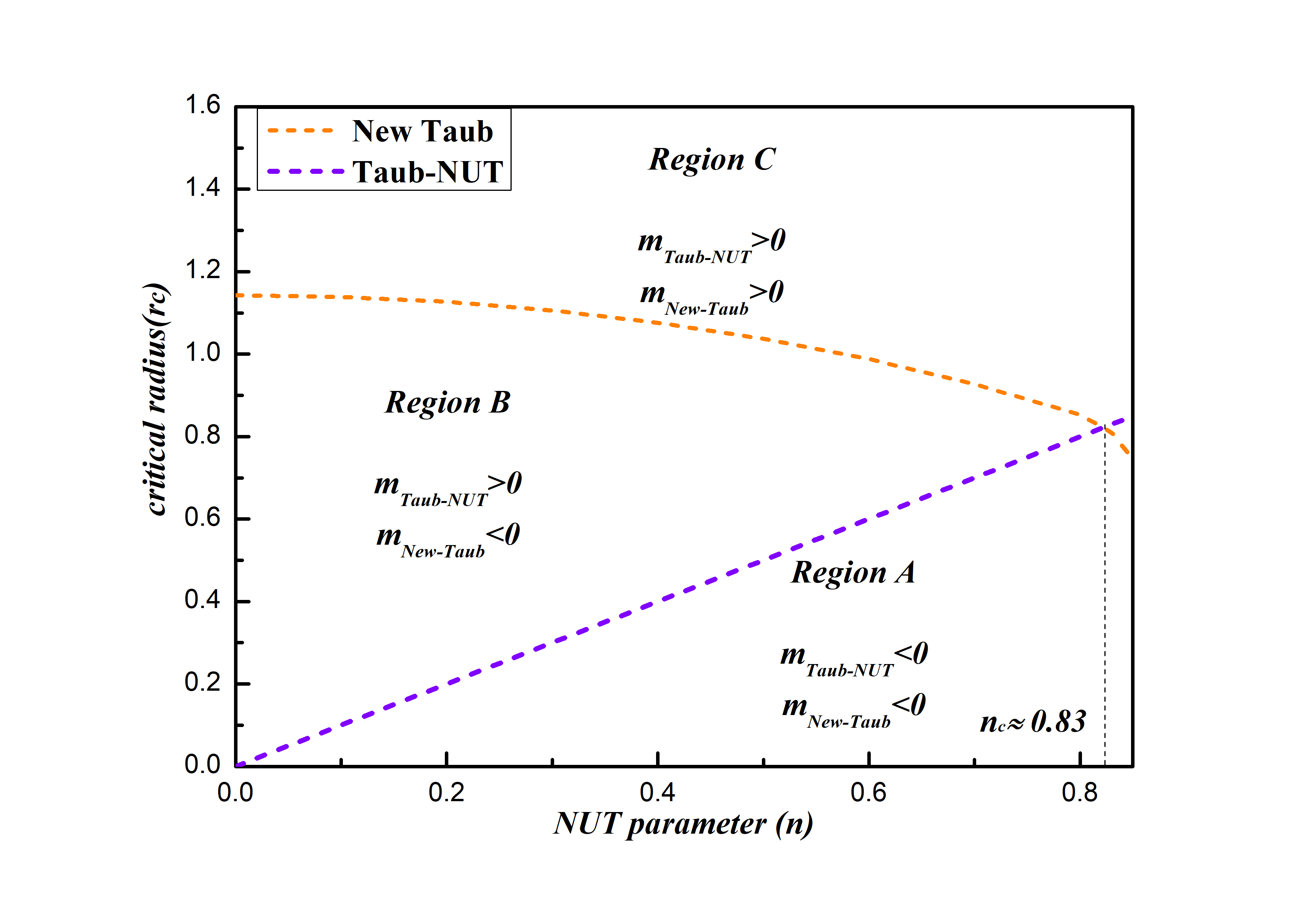}
    \caption{The $r_{0}-n$ curve of $m=0$ in Taub-NUT (violet line) and the new Taub (orange line). The three regions with different $m$ are shown as $Region~A,B,C$. The intersection of Taub-NUT and new Taub is tagged at $n\approx 0.83$.}
    \label{fig:3}
\end{figure}

We also plot the mass parameter as a function of horizon radius in Fig.5. The red line is the new found Taub-NUT like black holes, while the blue line represents the analytic Taub-NUT black hole. One can see that the mass parameter decreases as the event horizon $r_0$ increases. For $n=0.5$, the mass parameter can be negative when the black hole is larger enough. However, for larger n (like n=1.5), the mass parameter is always negative for the newly found Taub-NUT like black holes.  It is consistent with Fig.2, mass parameter decreases as NUT parameter grows. Compared with the Schwarzschild case in \cite{Lu:2015cqa}, our new Taub-NUT solutions have two different properties. The first is that the new solutions have two branches and the smooth connected. The other is that, for fixed NUT parameter n, there is a minimal value of event horizon radius,$r_{min} = 0.721$ for $n=0.5$ and $r_{min}=0.67$ for $n=1.5$.
\begin{figure}
    \centering
    \includegraphics[scale=0.24]{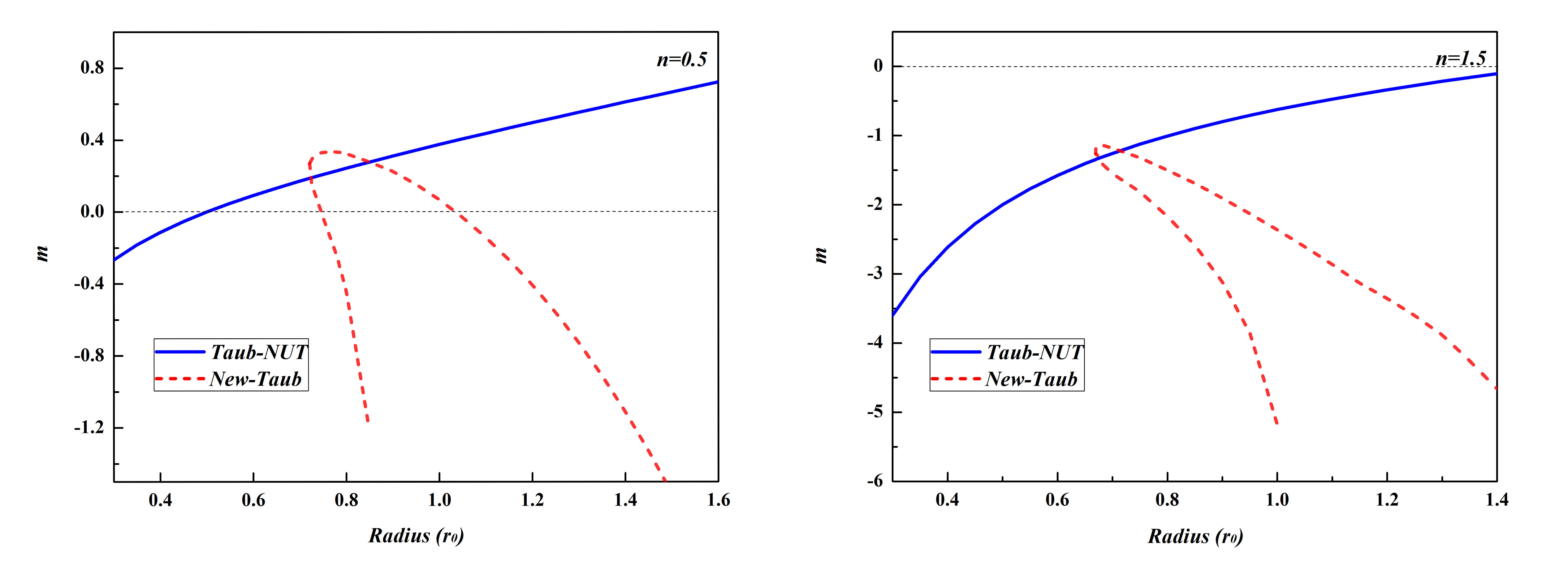}
    \caption{The $m-r_{0}$ curve of analytic Taub-NUT (blue line) and the new Taub-NUT black holes (red line) for different NUT parameter (n=0.5 for left plot and n=1.5 for right plot).  }
    \label{fig:4}
\end{figure}

We now turn to the thermodynamical properties of the newly found Taub-NUT black hole solution. The Hawking temperature can be calculated through the standard method
\begin{equation}
    T=\frac{\sqrt{f'(r_{0})h'(r_{0})}}{4\pi}=\frac{\sqrt{1+r_{0}\delta}}{4\pi r_{0}}\,.
\end{equation}

\begin{figure}
    \centering
    \includegraphics[scale=0.15]{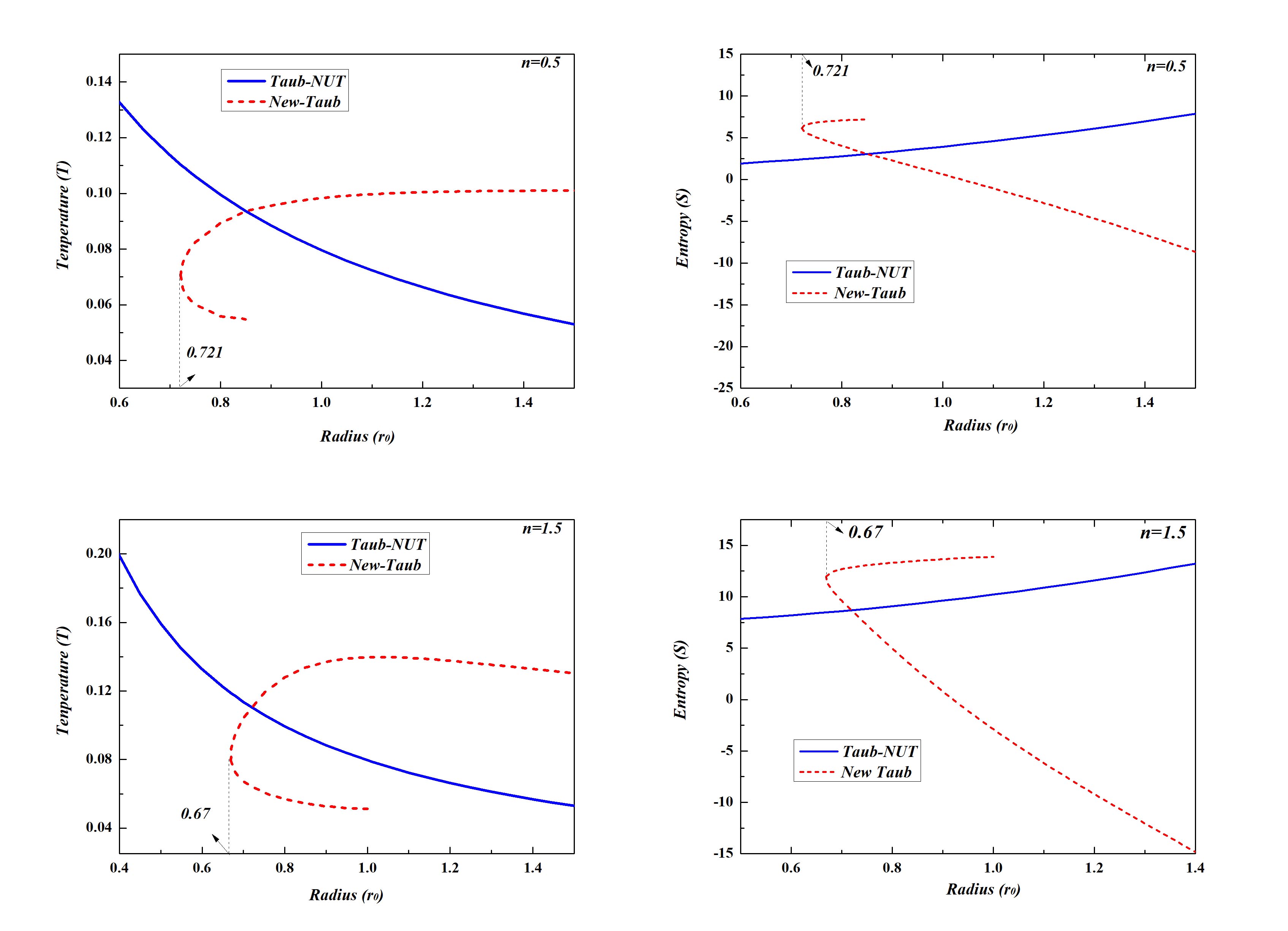}
    \caption{The Hawking temperature and Wald entropy in Taub-NUT and the new Taub-NUT solution with fixed $n=0.5$ are shown in the upleft and upright plots. And the homologous plots with $n=1.5$ are shown in the bottom. The Taub-NUT black holes is shown as blue full line, the first new Taub black holes are shown as red dash lines .}
    \label{fig:5}
\end{figure}

The temperature of the solution as a function of event horizon radius is shown in Fig.6 for $n=0.5$ and $n=1.5$. As a comparison, we also plot the temperature cure of Taub-NUT solution, the two curves cross at $r_0=0.852$ for $n=0.5$ and at $r_0=0.716$ for $n=1.5$. It is consistent with the Lichnerowicz mode analysis in the previous section. There is a maximum value of temperature, which is more obvious for $n=1.5$. There can exist two black holes in the same temperature. Furthermore, with $r_{0}$ decreasing, a notable inflection point will occur at $r_{0}=0.721$ in the $n=0.5$ case. As is seen in Fig.6, the two branches of the new solution are smoothly connected in the minimal radius. And the temperature of one branch is always higher than the other one.

The entropy of the newly found black hole is not equal to one quarter of the event horizon area due to the higher derivative terms. According to Wald's entropy formula \cite{Wald:1993nt,Iyer:1994ys}, the expression of entropy with higher curvature correction is
\begin{equation}
    S=-\frac{1}{8}\int \sqrt{h}d\Omega\epsilon_{ab}\epsilon_{cd}\frac{\partial L}{\partial R_{abcd}}=\pi(r_{0}^2+n^2)-4\pi \alpha r_{0} \delta  .
\end{equation}
As is mentioned in the beginning of this paper, the Gauss-Bonnet terms is a total derivative term in four dimensional spacetime and do not contribute to the equations of motion. However, the Gauss-Bonnet terms can have a constant contribution to the entropy. We use this freedom to make the entropy is equal to one quarter of the event horizon area for analytic Taub-NUT black hole solution. We plot the $S-r_0$ curve for $n=0.5$ and $n=1.5$ in Fig.6, the solid line is the cure for analytic Taub-NUT solution. Here, the two branches of the new solution meet at the minimal radius, and the entropy of one branch is always larger than the other branch. However, it is worth pointing out that the branch with lower temperature has larger entropy, and vice versa.

\textit{Conculsion.}- We constructed new Taub-NUT-like black hole solutions in Einstein gravity plus quadratic curvature terms. A Lichnerowicz mode analysis was carried out on the Ricci-flat Taub-NUT background, which gives us values of $(m,n)$ where there could exist new black holes carrying the massive spin-2 hair. Then new Taub-NUT like black hole solutions which carry spin-2 hair were constructed numerically, enlarging the black hole family in higher derivative gravity theories. The newly found solutions bifurcate from the Ricci-flat Taub-NUT solutions. Compared with the analytic Taub-NUT solutions, the temperature of one branch new solution has a maximum for a large $n$, and there exists two black holes with the same temperature. Since the Ricci-flat Taub-NUT is a solution, there can be a total of three black holes carrying NUT parameter for an appropriate given temperature.  This implies that there could be phase transitions between these black holes and the phase diagram becomes much richer.

As pointed out in \cite{Liu:2022wku} the mass of the analytic Taub-NUT black hole has extra contribution from the Minser strings which makes the total mass of the spacetime non-negative, though the mass parameter of the Taub-NUT solutions can be negative. The situation here is similar, the mass parameter can be negative, but negativity of the total mass could be avoided due to the extra contribution from the polar strings. However, the spin-2 modes are ghost like, the solution may still suffer from negative mass problem. The definition of mass is a difficult problem, but is very crucial at the same time, and so is the NUT charge. Both the definitions of mass and NUT charge are worth further exploring.

\textit{Acknowledgement.}- We thanks Ze Li and Hong Lu for useful discussion. We are also grateful for Ling-Yu Zhang's help in figures. This work is supported in part by NSFC (National Natural Science Foundation of China) Grants No.~12075166.

\appendix

\section{Full Analysis of Linechrowicz modes}

In section 3, it was shown that once the Lichnerowicz equation $\Delta_L \psi_{\mu\nu} = \lambda \psi_{\mu\nu}$, with eigenvalue $\lambda = -1/(2\alpha)$, has a a transverse traceless eigenfunction $\psi_{\mu\nu}$, there exists a linearized perturbation away from the Taub-NUT solution. Here, we want to give a comprehensive analysis of the Lichnerowicz equation.
 Since the perturbation is under  the Taub-NUT black hole background (\ref{taub}), we consider a symmetric tensor whose non-zero components are
\begin{equation}
\begin{split}
       \psi_{00}=h\psi_{0}(r),~\psi_{03}=\psi_{30}=2nh\cos\theta \psi_{0}(r) \,,
       \\ \psi_{11}=h^{-1}\psi_{1}(r),~\psi_{ij}=(r^2+n^2)\gamma_{ij}\psi_{2}(r),  \end{split}
\end{equation}
where $\psi_{0}$, $\psi_{1}$ and $\psi_{2}$ are unknown functions of radial coordinate $r$ only and $\gamma_{ij}$ are the induced metric of $d\Omega ^2$. Imposing the tracefree condition $g^{\mu\nu}\psi_{\mu\nu}=0$ and the transverse condition $\nabla^{\mu}\psi_{\mu\nu}=0$, we can obtain
\begin{equation}
    - \psi_0+\psi_1+2\psi_2 =0,
\end{equation}
and
\begin{equation}
    \psi_1' + \frac{h'}{2 h} (\psi_0+\psi_1) + \frac{ 2 r }{r^2+n^2} (\psi_1 - \psi_2) =0.
\end{equation}

With these equations, we can express $\psi_{0}$ and $\psi_{2}$ in terms of $\psi_{1}$ and $\psi_{1}'$. Then the $rr$ direction of the Lichnerowicz equations $\Delta_L \psi_{\mu\nu} = \lambda \psi_{\mu\nu}$ gives a equation of  second derivative of $\psi_1$
\begin{widetext}
\begin{equation}\label{leq}
    \begin{split}
        &\frac{2 \left(2 m^2 \left(7 n^2 r^2+n^4-6 r^4\right)+ m \left(-30 n^2 r^3+7 n^4 r+11 r^5\right)-6 n^4 r^2+13 n^2 r^4+3 n^6-2 r^6\right) \psi_1'}{\left(n^2+r^2\right)^2 \left(-m n^2+3 m r^2+3 n^2 r-r^3\right)}  \\
&+\frac{\left(r^2 - 2 m r - n^2\right) \psi_1''}{n^2+r^2}  + \left(\lambda - \frac{8 \left(4 m^2 n^2 r + m \left(-6 n^2 r^2 + n^4 + r^4 \right)+4 n^2 r^3\right)}{\left(n^2+r^2\right)^2 \left( m \left(n^2-3 r^2 \right)-3 n^2 r+r^3\right)} \right) \psi_1  = 0 .
    \end{split}
\end{equation}
\end{widetext}

In the large $r$, the equation turns out to be
\begin{equation}
    \psi_1 '' + \frac{4}{r} \psi_1' + \lambda \psi_1 = 0 \,.
\end{equation}

This equation can be easily solved and two Yukawa like modes,$\psi_1 ~ e^{\pm \sqrt {- \lambda r}}$, emerge. Generally, the full equation can not be solved analytically, and for numerical solution, one need to explore the appropriate values of the parameters to eliminate the diverging mode $e^{+ \sqrt {- \lambda r}}$ through the shooting method, and only the  normalisable mode $e^{ - \sqrt {- \lambda r}}$ survives.

To numerically solve the quation (\ref{leq}), we first do a near-horizon Taylor expansion for $\psi_1$ and then use this as initial data at the horizon to perform numerical integration out to large $r$. Without loss of generality, we set $\alpha = 1/2$, and find that there exists one mode corresponding to each value of $n$, and  the relation between the starting point  $r_0$ of the linear perturbation and $n$ are shown in Fig 1.
\end{document}